%% file: main.tex
\documentclass[mnsc,copyedit]{informs4}
\usepackage{eqndefns-left} 
\RequirePackage{newtxtext}
\RequirePackage{newtxmath}
\RequirePackage{bm}
\RequirePackage{endnotes}

\usepackage[T1]{fontenc}    
\usepackage{hyperref}       
\usepackage{url}            
\usepackage{booktabs}       
\usepackage{amsfonts}       
\usepackage{nicefrac}       
\usepackage{microtype}      
\usepackage{lipsum}		
\usepackage{doi}

\usepackage[ansinew]{inputenc} 
\usepackage[toc,page]{appendix}

\usepackage{algorithm}
\usepackage{algpseudocode}

\usepackage{mathtools}
\usepackage{pgfplots}
\usepackage{graphicx}
\usepackage{caption}
\usepackage{subcaption}
\pgfplotsset{compat = newest}

\usepackage{afterpage}
\usepackage{lscape} 
\usepackage{float}
\usepackage{booktabs}       
\usepackage{tikz}
\usepackage{pgfplots}
\usetikzlibrary{shapes.geometric, arrows, trees}
\usepackage{lipsum}

\usepackage{natbib}
 \bibpunct[, ]{(}{)}{,}{a}{}{,}%
 \def\bibfont{\small}%

\tikzstyle{nn_node} = [circle, 
minimum size=1cm, 
text centered, 
draw=black, 
fill=white!30]
\tikzstyle{arrow} = [thick,->,>=stealth, draw=gray]
\tikzstyle{annotation} = [rectangle, rounded corners, 
minimum width=1cm, 
minimum height=2cm,
text centered, 
draw=black, 
fill=lightgray]

\newsavebox{\Citname}

\newtheorem{hypothesis}{Hypothesis}
\newtheorem{rev_hypothesis}{Revised Hypothesis}

\hypersetup{
pdftitle={TBD},
pdfsubject={TBD},
pdfauthor={Matthias Lehmann},
pdfkeywords={TBD},
}

\begin{document}
%



\RUNAUTHOR{Lehmann, Cornelius, and Sting}

\RUNTITLE{AI Meets the Classroom}

\TITLE{AI Meets the Classroom: When Do Large Language Models Harm Learning?}

\ARTICLEAUTHORS{%
\AUTHOR{Matthias Lehmann}
\AFF{University of Cologne, \EMAIL{matthias.lehmann@wiso.uni-koeln.de}}

\AUTHOR{Philipp B. Cornelius}
\AFF{Rotterdam School of Management, Erasmus University, \EMAIL{cornelius@rsm.nl}}

\AUTHOR{Fabian J. Sting}
\AFF{University of Cologne, Rotterdam School of Management, Erasmus University, \EMAIL{sting@wiso.uni-koeln.de}}
} 

\ABSTRACT{%
The effect of large language models (LLMs) in education is debated: Previous research shows that LLMs can help as well as hurt learning. In two pre-registered and incentivized laboratory experiments, we find no effect of LLMs on overall learning outcomes. In exploratory analyses and a field study, we provide evidence that the effect of LLMs on learning outcomes depends on usage behavior. Students who substitute some of their learning activities with LLMs (e.g., by generating solutions to exercises) increase the volume of topics they can learn about but decrease their understanding of each topic. Students who complement their learning activities with LLMs (e.g., by asking for explanations) do not increase topic volume but do increase their understanding. We also observe that LLMs widen the gap between students with low and high prior knowledge. While LLMs show great potential to improve learning, their use must be tailored to the educational context and students' needs.
}%





\maketitle

\input{intro.tex}
\input{background.tex}

\input{theory.tex}
\input{studies_overview.tex}
\input{study_1.tex}
\input{study_2.tex}
\input{revised_hypotheses.tex}
\input{results.tex}
\input{discussion.tex}

\THEEndNotes



\bibliographystyle{informs2014}
\bibliography{refs}  






\newpage

\begin{appendices}

\input{screenshots.tex}
\input{covariates.tex}

\input{message_coding.tex}
\input{robustness.tex}

\end{appendices}

\end{document}

%% file: intro.tex
\section{Introduction}

Large language models (LLMs) have become pervasive in education. They can reproduce knowledge on a vast range of subjects and perform well above passing grades in university-level exams \citep[e.g.,][]{drori2022neural, terwiesch2023would}. As a result, the majority of students now use LLMs regularly \citep{freeman2025student}. As business adoption of LLMs spreads throughout the economy, emerging research points to significant productivity gains \citep{brynjolfsson2023generative, dell2023navigating, noy2023experimental}. Yet, despite similar if not greater adoption of LLMs in education, their impact on learning outcomes remains contentious. Whereas some studies show that LLMs help learning \citep{nie2024gpt, kestin2024ai}, others find the opposite \citep{bastani2024generative}.

To make a step toward reconciling these conflicting findings on whether LLMs help or harm learning, we report on a field study and two pre-registered and incentivized laboratory experiments. In the two experiments, we find no significant effect of LLM access on overall learning outcomes. In subsequent exploratory analyses, we provide evidence that the effect depends on how students use LLMs and how learning outcomes are measured. Students who use LLMs to substitute some of their learning activities (e.g., practice exercises) move faster through the instructional material and thus increase the volume of topics they can cover. However, this increase in speed comes at the cost of a shallower understanding of each topic. In contrast, students who use LLMs to complement their learning activities (e.g., using LLMs as tutors) improve their topic understanding. In the field study, in which we only observe substitutive use, we leverage exogenous variation in LLM availability to find a significant long-term decline in overall learning outcomes. In addition, we show that how LLMs affect learning is contingent on students' prior knowledge: LLMs support the learning of students with more prior knowledge but harm the learning of students with less prior knowledge. Unlike in workplaces \citep{brynjolfsson2023generative, dell2023navigating}, LLMs appear to increase inequality in education.

We make several empirical and theoretical contributions to the literature on education operations \citep[e.g.,][]{smilowitz2020use, yoo2023economic, keppler2022crowdfunding, keppler2024little} and education technology \citep[e.g.,][]{zhang2017does, corsten2023get, bray2024tutorial}. First, we study how unrestricted LLMs affect learning outcomes. Previous studies restricted LLMs to certain use cases, in particular by limiting substitutive use \citep{nie2024gpt, kestin2024ai}. While such customized LLMs offer great opportunities in supervised educational settings, a large part of learning occurs unsupervised, and during such unsupervised learning most students have access to free and unrestricted LLMs, such as ChatGPT. Second, we are among the first to investigate the effect of LLMs on learning in a laboratory without potentially confounding effects from undocumented LLM use (e.g., at home), classrooms, teachers, and peers. We hope that our simple experimental setup provides a basis for more laboratory studies on the microfoundations of learning with LLMs. Third, we add to the emerging literature on how technology use influences learning \citep[e.g.,][]{teodoridis2018understanding, wuttke2022seeing}. Our results suggest that students prefer to use LLMs to substitute rather than complement their own learning activities, which can harm learning outcomes. Despite an increasing focus on student behavior \citep{lavecchia2016behavioral, damgaard2020behavioral}, issues such as user self-control have been largely absent from the evaluation of technology in education \citep{bulman2016technology, chatterji2018innovation}. Our study also offers insights to broader innovation scholarship about the impact that technology use has on learning and knowledge acquisition \citep[e.g.,][]{shane200450th, loch2017creativity}.

The rest of the paper is organized as follows. In Section \ref{sec:literature}, we review the literature and in Section \ref{sec:theory} we develop hypotheses. We provide an overview of all studies in Section \ref{sec:overview_studies} and present the field evidence in Section \ref{sec:s1}. Section \ref{sec:s2} presents the first experiment and Section \ref{sec:revised_hypotheses} our revised hypotheses based on that experiment. Section \ref{sec:s3} presents the second experiment and Section \ref{sec:exploratory} our exploratory analyses. We discuss our findings in Section \ref{sec:discussion}.

%% file: background.tex
\section{Related Literature}\label{sec:literature}

Evidence on how technology affects learning outcomes remains mixed \citep[for reviews, see][]{bulman2016technology, chatterji2018innovation}. On the one hand, self-paced and individualized computer-aided instruction promises to extend the benefits of small class sizes and one-on-one interaction with instructors to more students. In addition, technology also supports traditional learning activities, for example by allowing instructors to closely track students' learning progress or by making instruction more engaging. On the other hand, technology may be distracting, because it draws students' attention away from the subject matter and toward understanding the technology itself. Technology that reduces job demands in work settings is sometimes found to reduce learning \citep{wang2020does} and education technology that similarly lowers effort may thus also reduce learning. Moreover, investment in technology may take away from investment in traditional learning activities. Despite the importance of technology in education, rigorous evaluations of its efficacy remain scarce.

Large language models (LLMs) have demonstrated astounding capabilities in educational assessments \citep[e.g.,][]{drori2022neural, terwiesch2023would}. Teachers who collaboratively ``think'' with LLMs about teaching activities report being more productive when preparing classes, while teachers who let LLMs generate teaching material report no change in their productivity \citep{keppler2024backwards}. For students, however, the emerging research has found mixed results. Students perceive LLM tutors to be helpful \citep{liu2024teaching}, but prefer human explanations over LLM explanations \citep{pardos2023learning, prihar2023comparing}. Still, LLM explanations improve students' learning outcomes \citep{kumar2023math, pardos2023learning}.

In the above studies, students were shown LLM-generated content, but they could not interact with an LLM themselves. \citet{bastani2024generative} run a field experiment in math classes at a high school in which students interact with an LLM. They find that the LLM improves student performance on practice problems but reduces performance in the final exam when students no longer have access to it. In addition, they show that this harmful effect on learning can be mitigated by modifying the LLM so that it produces fewer errors and does not directly provide solutions to students. \citet{nie2024gpt} conduct a similar field experiment in an online programming course. They observe very limited usage of the LLM and document that it reduces engagement with the course materials and the likelihood of completing the course. Still, the small group of students who do use the LLM performs better on the final exam. \citet{kestin2024ai} conduct a field experiment in university-level physics classes, finding that students achieve higher learning outcomes when learning with an LLM-based personal tutor than when attending traditional active learning classes. \citeauthor{nie2024gpt} and \citeauthor{kestin2024ai} also implement safeguards against students asking an LLM for solutions to practice problems. Given these contrasting findings, we extend the nascent literature on LLMs in education by shedding light on the circumstances under which LLMs help or hurt learning.

%% file: theory.tex
\section{Theory}\label{sec:theory}

We conceptualize learning outcomes as the value added by education, which is the difference in students' knowledge before and after education \citep{hanushek2020education}. In a simple framework of a learning process, we decompose learning outcomes into topic volume and topic understanding.  Students learn about a set of topics, such as the topics taught in a course. To learn about each topic, students engage in a variety of learning activities, for example reading textual explanations or solving practice problems. Students learn as they cover more topics and understand each topic better. Thus, students' learning outcomes are a function of topic volume and understanding.\endnote{The distinction between topic volume and topic understanding bears some resemblance to the distinction between education quantity and quality in economics \citep{hoekstra2020returns}, but it is not exactly the same. Education quantity usually refers to the number of years people spend in education, while education quality refers to the efficacy of educational institutions in imparting knowledge. In our case, the distinction is similarly about the volume of knowledge that students are exposed to and the efficacy of internalizing that knowledge, but the level of analysis is at the individual learning process rather than at the institution.} Based on this framework, we develop two competing and pre-registered\endnote{\url{https://aspredicted.org/2LM_DP3}\label{endnote:prereg1}} hypotheses of the effect of LLMs on learning outcomes.

\subsection{LLMs Improve Learning Outcomes}\label{sec:pro_llm}

Personal tutoring improves student learning outcomes \citep{cohen1982educational, bowman2013academic}. LLMs promise to provide personal tutoring at scale, to every student, at all hours of the day. Prior studies have found that even previous rule-based, non-LLM chatbots improve learning \citep{abbasi2014measuring, chen2020chatbot, yin2021conversation,  chang2022promoting, essel2022impact, ait2023impact}. LLMs improve on rule-based chatbots in several ways.

First, the large-scale pretraining of LLMs equips them with vast knowledge across a wide range of domains \citep{brown2020language, touvron2023llama}. LLMs also exhibit strong coding abilities, which includes problem-solving and the ability to reason about self or user-generated code \citep{touvron2023llama, team2023gemini, liu2024your}. In this domain, LLMs have surpassed the average  performance of human professionals \citep{team2023alphacode}.

Second, modern LLMs are fine-tuned to act as helpful assistants \citep{openai2024most, touvron2023llama}. This allows students to interact with them flexibly and iteratively like they would with a human tutor. LLMs understand user queries even if they do not use the correct technical jargon and tailor their responses to the level of knowledge that the user exhibits in their message. The versatility of LLMs means that they can give hints on problems, provide examples, explain concepts, and write code.

Third, LLMs can act as a thought partner for students and enhance understanding by providing detailed explanations and step-by-step solutions \citep{mollick2022new, kasneci2023chatgpt, meyer2023chatgpt, extance2023chatgpt}. Several studies have found that LLM-generated explanations improve learning in mathematics \citep{pardos2023learning, prihar2023comparing, kumar2023math} and physics \citep{kestin2024ai}. Similarly, learning improved in programming courses in which students had access to LLMs \citep{liu2024teaching, nie2024gpt}.

Interacting with LLMs  while learning may affect both components of learning: volume and understanding. LLMs may allow students to cover more topics, because students can quickly ask questions if they do not understand something. For the same reason, LLMs may also improve students' understanding of each topic. For instance, an LLM may explain a topic that a student has not understood before. Thus, LLMs may improve learning outcomes through both topic volume and topic understanding.

\begin{hypothesis}\label{hyp:hypo1}
	Having access to LLMs when learning to code increases the learning outcome of students.
\end{hypothesis}

\subsection{LLMs Harm Learning Outcomes}\label{sec:contra_llm}

Learning and application often go hand in hand: To fully understand a topic, students need to apply their knowledge to a related problem and learn from their experience of solving that problem. Such learning-by-doing is especially important when learning to code \citep{narayanan2009matter}. As we mentioned above, LLMs can generate working code and students may be tempted to use this functionality to solve practice exercises more quickly. If students use an LLM to solve exercises (fully or partially), they reduce their own mental effort and do not engage in the problem solving necessary to internalize and understand a topic. Hence, LLM usage may inhibit learning-by-doing and thus decrease learning outcomes.

This argument extends beyond learning-by-doing, because LLMs may reduce time and effort spent on learning more broadly. Successful learning requires investments of time and effort and students are easily lured into reducing their investments if given the opportunity \citep{bishop2006drinking, lavecchia2016behavioral}. For example, by relying on LLMs to talk about and explain concepts, students may inadvertently reduce their own mental effort during the acquisition of new knowledge. Previous technologies from computers, to virtual reality, to pre-LLM artificial intelligence have similarly reduced the mental effort required of their users and thereby decreased their learning (e.g., \citealp{mueller2014pen}, \citealp{wuttke2022seeing}, \citealp{dell2022falling}, \citealp{miola2024gps}; for a review, see \citealp{wang2020does}).

Aside from reducing mental effort, LLMs may also harm learning by producing falsehoods or "hallucinations" \citep{kasneci2023chatgpt, meyer2023chatgpt, extance2023chatgpt}. While experts may be able to spot such errors, students are easily deceived because they do not know the topic yet. Once put on the wrong path, students may fundamentally misunderstand a new topic. In addition, LLMs may engage students in tangential or peripheral conversations that reduce the volume of topics students have time to cover. Thus, LLMs may harm learning outcomes and \citet{bastani2024generative} provide first empirical evidence that high school students learning mathematics with the help of LLMs perform worse in exams.

\begin{hypothesis}\label{hyp:hypo2}
	Having access to LLMs when learning to code decreases the learning outcome of students.
\end{hypothesis}

%% file: studies_overview.tex
\section{Overview of Studies}\label{sec:overview_studies}

We test the hypothesized effects of LLMs on learning in three studies. We combine field data (Study 1) and two lab experiments (Studies 2 and 3) to provide evidence of both internal and external validity \citep[e.g.][]{lee2021managing, buell2021last, wiltermuth2023anchoring}. In Study 1, we show how the use of LLMs during weekly assignments affects learning in two graduate programming courses. In Studies 2 and 3, we replicate the field setting in two incentivized and pre-registered laboratory experiments. In Study 2, we manipulate LLM access and compare learning outcomes between treated (with LLM) and controls (without LLM). Study 3 is an exact replication of Study 2 except for the availability of copy and paste for all participants (treated and controls). In Study 2, participants could not copy and paste text from or to the LLM and---as we discuss in Section \ref{sec:revised_hypotheses}---this may have affected how the LLM was used. In the next sections, we describe each study and report its results.

%% file: study_1.tex
\section{Study 1: Field Evidence}\label{sec:s1}

We begin with a field study of two graduate programming courses delivered at roughly the same time in the Spring of 2023 at a public Dutch university. The courses started with an introduction to Python programming (e.g., variables, data types) and finished with simple machine learning models (e.g., random forests). The two courses were delivered in different programs, one in information systems (IS) and the other in business analytics (BA). The courses were identical in terms of content and assessment except that the BA course had one less lecture on Python because students in that program had already received an introduction to Python in another course. Hence, the IS course comprised five lectures over five weeks and the BA course comprised four lectures over four weeks. 56 students took the IS course, and 57 students took the BA course; no student took both courses.

Each lecture was accompanied by a coding homework assignment on an online coding platform. Students had three days to complete each assignment and their assignment grade counted towards their course grade. Each assignment consisted of a different number of coding questions and in total there were 65 questions across five assignments in the IS course and 55 questions across four assignments in the BA course. (There is one less assignment in the BA course because there is also one less lecture.) On the online coding platform, students could submit code for each question as often as they liked and for each submission they immediately saw their grade and any potential error message that their code generated. Only their last code submission was graded. While the lecture imparted programming theory, students had to work through the assignments to fully internalize the content. Thus, students were learning by doing, that is by working through the assignments and receiving feedback in form of their grade and potential error messages.

The data has a panel structure and the unit of analysis is student--questions. There are 56 $\times$ 65 = 3,640 student--question observations in the IS course and 57 $\times$ 55 = 3,135 student--question observations in the BA course, yielding 6,775 student--question observations in total.

\subsection{Variables}

\subsubsection{Dependent Variable}

Students' grade for each question is the dependent variable. The grade is normalized such that for student $i$ and question $q$, $\mbox{Grade}_{iq} \in [0,1]$. At the focal institution, the average normalized course grade is between 0.7 and 0.8. An average (across all courses) of at least 0.825 is considered a distinction and an average of at least 0.9 is considered an exceptional distinction. A grade below 0.55 is considered a fail. Most students fall between 0.6 and 0.85.

\subsubsection{Explanatory Variable}

We measure students' use of generative AI as the similarity between students' final code submission and ChatGPT-generated code for the same question. ChatGPT was the main LLM chatbot available during the courses. We collected ChatGPT-generated code by submitting the same question descriptions that students saw to the OpenAI API. We exactly replicate the ChatGPT environment that was available to students at the time of the courses (\verb|gpt-3.5-turbo-0613|). Because ChatGPT generates solutions stochastically, which can therefore differ across queries, and we do not know which solution a student received, we query ChatGPT 50 times per question and take the maximum similarity across the 50 ChatGPT solutions:
\begin{align} \label{eq:fe}
& \mbox{ChatGPT Similarity}_{iq} = \notag \\
& \quad \max \Bigl\{ \mbox{sim}\bigl(\mbox{Code}_{iq}, \mbox{Code}^{\text{LLM}}_{jq}\bigr) \; \Big| \; j=1,\ldots,50 \Bigr\}  \mbox{,} \notag
\end{align}
for student $i$ and question $q$, where $\mbox{Code}_{iq}$ is the final student code, $\mbox{Code}^{\text{LLM}}_{jq}$ is one of the 50 ChatGPT generated solutions, and $\mbox{sim}(\cdot, \cdot) \in [0,1]$ is the normalized Damerau--Levenshtein similarity (available from RapidFuzz). In addition to \textit{ChatGPT Similarity} for a focal question, we compute the \textit{Cumulative ChatGPT Similarity} over all questions answered before the focal question to estimate the effect of past LLM usage on learning.

Since we cannot directly observe whether a student used ChatGPT, \textit{ChatGPT Similarity} is a proxy for this behavior: A higher value maps into a higher probability that ChatGPT was used. This measurement of treatment is similar to intention-to-treat analyses in experimental designs, in which treatment assignment as the main explanatory variable also corresponds to a higher probability of actual treatment \citep[e.g.,][]{bastani2024generative}.

We exclude 181 observations of questions that students did not attempt to answer. For such unanswered questions, we cannot measure \textit{ChatGPT Similarity}, because no student code was submitted. Unanswered questions are conceptually different from answered questions with a true \textit{ChatGPT Similarity} of 0, so imputing a \textit{ChatGPT Similarity} of 0 for unanswered questions would introduce a measurement bias.

\subsubsection{Control Variables}

We control for the following potential confounders:

\textit{Plagiarism}. Because the homework assignments are not invigilated, students may copy solutions from one another. If a student whose solution was plagiarized used ChatGPT, the plagiarizing student's \textit{ChatGPT Similarity} will also be higher even if they themselves did not use ChatGPT. For student $i$ and question $q$ we measure $\mbox{Plagiarism}_{iq}$ as the maximum normalized Damerau--Levenshtein similarity with any other student solution for the same question in the same course. We also compute \textit{Cumulative Plagiarism} over all questions answered before the focal question.

\textit{Questions Answered}. Students' performance and behavior may change as they solve more questions. For example, they may become more confident and use ChatGPT less. We thus control for the cumulative number of questions students have answered before the focal question.

\textit{Previous Grade}. Students' performance and behavior may change depending on their performance in previous questions. For example, students who are doing well may become more confident and use less ChatGPT. We thus control for the average grade across all previously answered questions.

\begin{table}
    \TABLE
    {Study 1: Descriptive Statistics.\label{tab:s1_desc_stats}}
	{\begin{tabular}{lcccc}
		\toprule
		 & Mean & Min. & Max. & S.D. \\
		\midrule
		\textit{Grade} & 0.95 & 0.00 & 1.00 & 0.19 \\
		\textit{ChatGPT Similarity} & 0.42 & 0.00 & 0.99 & 0.14 \\
		\textit{Cum. ChatGPT Similarity} & 12.19 & 0.00 & 37.87 & 7.93 \\
		\textit{Plagiarism} & 0.71 & 0.00 & 1.00 & 0.19 \\
		\textit{Cum. Plagiarism} & 22.15 & 0.00 & 54.58 & 12.89 \\
		\textit{Questions Answered} & 29.07 & 0.00 & 64.00 & 17.45 \\
		\textit{Previous Grade} & 0.96 & 0.00 & 1.00 & 0.15 \\
		\textit{Time Taken} & 47.08 & 0.10 & 1,595.7 & 103.02 \\
		\textit{Cum. Time Taken} & 795.72 & 0.00 & 7,153.07 & 900.25 \\
		\textit{ChatGPT Outages} & 882.21 & 0.00 & 58,500.00 & 5,735.40 \\
		\textit{Cum. ChatGPT Outages} & 14,748.96 & 0.00 & 827,092.00 & 75318.07 \\
		\bottomrule
	\end{tabular}}
    {\textit{Note}. N = 6,594.}
\end{table}

\subsection{Identification}

Our baseline model is a two-way fixed effects (FE) model. For student $i$ and question $q$,
\begin{align}
\mbox{Grade}_{iq} &= \beta_1 \mbox{ChatGPT Similarity}_{iq} \notag \\
& \quad + \beta_2 \mbox{Cum. ChatGPT Similarity}_{iq} \notag \\
& \quad + \bm{X'_{iq}} \bm{\beta_3} + c_i + c_q + \varepsilon_{iq} \mbox{,}
\end{align}
where $\bm{X}_{iq}$ is a vector of control variables, $c_i$ are student fixed effects, and $c_q$ are question fixed effects. This setup controls for constant additive student and question confounders as well as for dynamic linear confounders in the form of students' prior performance. The parameter $\beta_2$ captures the effect of past LLM use on the grade of the current question. It estimates how past LLM use has affected students' overall learning (both in terms of topic volume and understanding). Since we can only observe whether students submitted LLM-generated solutions and not whether they used the LLM in any other way that supported their learning, $\beta_2$ only estimates the effect of students substituting some of their learning activities with the LLM.

In addition and to rule out further confounding factors (e.g., measurement error), we estimate the FE model in Equation \ref{eq:fe} with instrumental variables for the two similarity measures. As instruments, we use ChatGPT service interruptions as reported on the OpenAI Status Blog\endnote{https://status.openai.com} (OpenAI is the vendor of ChatGPT). We include all interruptions that mention a ChatGPT outage in the incident report or as an affected service. We match each interruption to students who were solving a question at the same time, that is to all student--questions whose first submission occurred before the end of an interruption and whose last submission occurred after the start of an interruption. \textit{ChatGPT Outages} is then the amount of time (in seconds) that an interruption overlaps with the time between the first and last submission. \textit{Cum. ChatGPT Outages} is the corresponding cumulative measure over all previous questions. For instance, if a student worked on a question between 9am and 11am, and there was an outage from 9.30am until 10am, then for that student--question we calculate 30 minutes of outage. As students who take longer to solve a question may be exposed to more outages, we control for the time between first and last submission to a question as the \textit{(Cumulative) Time Taken} (in minutes). \textit{ChatGPT Outages} exogenously vary the availability of ChatGPT and thus allows us to estimate quasi-experimental effect sizes.

We estimate the fixed effects 2SLS (FE2SLS) estimator \citep{wooldridge_econometric_2010}, implemented as \verb|ivreghdfe| in Stata \citep{correia_linear_2017}. In the first stages, we estimate
\begin{align} \label{eq:fe2sls1}
& \mbox{ChatGPT Similarity}_{iq} \notag \\
& \quad = \gamma_1 \mbox{ChatGPT Outages}_{iq} \notag \\
& \quad \quad + \gamma_2 \mbox{Cum. ChatGPT Outages}_{iq} \notag \\
& \quad \quad + \bm{X'_{iq}} \bm{\gamma_3} + c_i + c_q + u_{iq}
\end{align}
and
\begin{align} \label{eq:fe2sls2}
& \mbox{Cum. ChatGPT Similarity}_{iq} \notag \\
& \quad = \delta_1 \mbox{ChatGPT Outages}_{iq} \notag \\
& \quad \quad + \delta_2 \mbox{Cum. ChatGPT Outages}_{iq} \notag \\
& \quad \quad + \bm{X'_{iq}} \bm{\delta_3} + c_i + c_q + v_{iq} \mbox{.}
\end{align}
In the second stage, we estimate
\begin{align} \label{eq:fe2sls3}
& \mbox{Grade}_{iq} \notag \\
& \quad = \beta_1 \widehat{\mbox{ChatGPT Similarity}}_{iq} \notag \\
& \quad \quad + \beta_2 \widehat{\mbox{Cum. ChatGPT Similarity}}_{iq} \notag \\
& \quad \quad + \bm{X'_{iq}} \bm{ \beta_3} + c_i + c_q + \varepsilon_{iq} \mbox{.}
\end{align}
As before, $\beta_2$ is the parameter of interest and estimates the effect of substitutive LLM use on overall learning outcomes.

\subsection{Results}

\begin{table}
    \TABLE
    {Study 1: Two-way Fixed Effects (FE) Results.\label{tab:s1_fixed_effects}}
	{\begin{tabular*}{250pt}{@{\extracolsep{\fill}}lc}
		\toprule
		 & \textit{Grade} \\
		\midrule
		\textit{ChatGPT Similarity} & 0.18*** (0.05) \\
		\textit{Cum. ChatGPT Similarity} & -0.02*** (0.00) \\
		\textit{Plagiarism} & 0.15*** (0.03) \\
		\textit{Cum. Plagiarism} & 0.00 (0.00) \\
		\textit{Questions Answered} & 0.00 (0.00) \\
		\textit{Previous Grade} & 0.04 (0.04) \\
		\textit{Time Taken} & 0.00* (0.00) \\
		\textit{Cum. Time Taken} & -0.00 (0.00) \\
        \midrule \midrule
        Constant & 0.79*** (0.05) \\
        \textit{F}-statistic & 7.64*** \\
        $R^2$ & 0.29 \\
        Adjusted $R^2$ & 0.27 \\
        Students & 113 \\
        Observations & 6,594 \\
		\bottomrule
	\end{tabular*}}
    {\textit{Notes}. Student-clustered standard errors in parentheses. Asterisks indicate significance: *$p$ <0.10; **$p$ <0.05; ***$p$ <0.01.}
\end{table}

\autoref{tab:s1_desc_stats} presents descriptive statistics of the panel data. The reported average grade is larger than the actual course grade, because we do not apply question weights in the statistical analyses. \autoref{tab:s1_fixed_effects} shows the results of the two-way fixed effects model in Equation \ref{eq:fe} (FE). The effect of \textit{ChatGPT Similarity} on the \textit{Grade} of the current question is positive and significant ($p$ < 0.001), while the effect of \textit{Cum. ChatGPT Similarity} is negative and significant ($p$ < 0.001). If students submit code that is identical to a ChatGPT-generated solution, their grade for the current question increases by 0.18, but their grade on all subsequent questions decreases by 0.02.

\begin{table}
    \TABLE
    {Study 1: Instrumental Variable Fixed Effects (FE2SLS) Results.\label{tab:s1_instruments}}
	{\begin{tabular}{lccc}
		\toprule
        & \multicolumn{2}{c}{First Stages} & Second Stage \\
		 & \textit{ChatGPT Similarity} & \textit{Cum. ChatGPT Similarity} & \textit{Grade} \\
		\midrule
        \textit{ChatGPT Outages} & -0.00*** (0.00) & -0.00 (0.00) &   \\
		\textit{Cum. ChatGPT Outages} & -0.00 (0.00) & -0.00*** (0.00) &   \\
		\textit{ChatGPT Similarity} &  &  & -0.35 (0.56)  \\
		\textit{Cum. ChatGPT Similarity} &  &  & -0.06*** (0.02)  \\
		\textit{Plagiarism} & 0.16*** (0.02) & 0.64*** (0.24) & 0.26** (0.10)  \\
		\textit{Cum. Plagiarism} & 0.00** (0.00) & 0.13** (0.06) & 0.01** (0.01)  \\
		\textit{Questions Answered} & -0.00 (0.00) & 0.40*** (0.06) & 0.02*** (0.01)  \\
		\textit{Previous Grade} & 0.02 (0.02) & -0.27 (0.35) & 0.04 (0.04)  \\
		\textit{Time Taken} & 0.00 (0.00) & -0.00 (0.00) & 0.00 (0.00)  \\
		\textit{Cum. Time Taken} & -0.00* (0.00) & -0.00*** (0.00) & -0.00** (0.00)  \\
        \midrule \midrule
        \textit{F}-statistic& 13.91*** & 5.60*** & 4.74*** \\
        Students & 113 & 113 & 113 \\
        Observations & 6,594 & 6,594 & 6,594 \\
		\bottomrule
	\end{tabular}}
    {\textit{Notes}. Student-clustered standard errors in parentheses. The constant term is partialled out. Asterisks indicate significance: *$p$ <0.10; **$p$ <0.05; ***$p$ <0.01.}
\end{table}

\autoref{tab:s1_instruments} shows the results of the instrumental variable fixed effects model in Equations \ref{eq:fe2sls1}--\ref{eq:fe2sls3} (FE2SLS). In the first stages, both instruments significantly affect the instrumented variables in the expected directions: \textit{ChatGPT Outages} reduces \textit{ChatGPT Similarity} ($p$ = 0.005) and \textit{Cum. ChatGPT Outages} reduces \textit{Cum. ChatGPT Similarity} ($p$ = 0.002). In the second stage, the effect of \textit{ChatGPT Similarity} on the \textit{Grade} of the current question is not significant anymore ($p$ = 0.334), while the effect of \textit{Cum. ChatGPT Similarity} remains negative and significant ($p$ = 0.002). If students submit code that is identical to a ChatGPT-generated solution, their grade for the current question does not change, but their grade on all subsequent questions decreases by 0.06.

Across the FE and FE2SLS models we find evidence in support of a negative effect of LLMs on learning outcomes. The more students substitute for their own learning by using LLMs to generate solutions, the less they learn and the worse they perform on subsequent questions. The benefit of using an LLM to get a better grade on the current question is only significant in the FE model, but not in the FE2SLS model. One explanation may be the correlation between high \textit{ChatGPT Similarity} and a correct solution, such that conditional on student ability ($c_i$) and prior performance (\textit{Previous Grade}), high \textit{ChatGPT Similarity} indicates a chance correct solution rather than LLM use. This is a form of measurement error and by using an instrumental variable, we remove the correlation between \textit{ChatGPT Similarity} and a correct solution and the estimate hence turns not significant.

%% file: study_2.tex
\section{Study 2: Laboratory Experiment}\label{sec:s2}

\input{methods.tex}

\subsection{Results}
We recruited 108 subjects, all of which completed the experiment. One subject solved the final question in the post-test and was thus excluded according to our pre-registered filter criteria. Of the remaining 107 subjects, almost everyone had either no prior coding experience in Python (79\%) or was a beginner (17\%). On average, subjects took 89 minutes to complete the experiment and earned 18.50 euros. \autoref{tab:covar_stats} in \autoref{sec:covariates} provides summary statistics of all pre-treatment variables.

\begin{table}
    \TABLE
    {Study 2: Correctly Answered Questions per Experimental Phase.\label{tab:stud2_overview}}
	{\begin{tabular}{lcccc}
		\toprule
		     & \textit{Pre-test} & \textit{Learning Phase}  & \textit{Post-test} & \textit{Post-test} $-$ \textit{Pre-test} \\
		\midrule
		Control Condition    & 3.1 (3.0)  & 15.1 (3.8) & 7.9 (4.2) & 4.8 (3.3) \\
		Treatment Condition    & 3.6 (3.5) & 16.3 (4.8)  & 9.0 (4.7) & 5.4 (3.5) \\
		\bottomrule
	\end{tabular}}
    {\textit{Note}. Standard deviations are in parenthesis.}
\end{table}
Subjects in the treatment condition sent 4.45 (median = 3, S.D. = 3.94) messages on average to the LLM and six subjects did not use the LLM at all.\endnote{The six subjects who did not use the LLM performed better than other treated (\textit{Post-test} $-$ \textit{Pre-test} = 7.8). Results are similar if we exclude these six subjects (see \autoref{tab:rob_posttest_treatment_excl_untreated} in \autoref{sec:robustness})} \autoref{tab:stud2_overview} shows the average performance of the treatment and control conditions in each of the three experimental phases. The treatment condition outperformed the control condition in each of the three phases, solving half a question more in the \textit{Pre-test} and one question more in both the \textit{Learning Phase} and the \textit{Post-test}. None of the three differences are statistically significant in two-tailed Welch's \textit{t}-tests ($p$ = 0.425, $p$ = 0.175, $p$ = 0.206). The difference of \textit{Post-test} $-$ \textit{Pre-test}, which adjusts for differences in prior coding experience, is half a question larger in the treatment condition, but also not significant ($p$ = 0.376).

In addition, we report the results of regressions in \autoref{tab:reg_s2_main}, in which we adjust for observable pre-treatment characteristics to control for random differences between the control and treatment conditions. As in the model-free analysis, the treatment effect is not significant (column 1, $p$ = 0.580).

In our theoretical framework, LLMs can affect learning outcomes by changing the volume of topics subjects can study and/or by enhancing or harming their understanding of each topic. While we cannot observe whether subjects have understood a topic directly, we can estimate the effect of the LLM on understanding by controlling for the LLM's effect on the volume of topics covered during the \textit{Learning Phase}. Any remaining treatment effect must be due to subjects' increased understanding. In the second column of  \autoref{tab:reg_s2_main}, we thus control for subjects' progress during the \textit{Learning Phase}. In this setup, too, the treatment effect remains not significant ($p$ = 0.562). In the third column, we can see that LLMs also do not affect volume ($p$ = 0.214). Hence, we do not find support for a positive or a negative effect of LLMs on learning.

\begin{table}
    \TABLE
    {Study 2: Regression Analyses.\label{tab:reg_s2_main}}
{\begin{tabular}{lccc}
\toprule
& \multicolumn{2}{c}{\textit{Post-test}} & \textit{Learning Phase} \\
& (1) & (2) & (3) \\
\midrule
\textit{Treatment} (LLM access = 1) & 0.396 (0.713) & -0.276 (0.474) & 0.828 (0.662) \\
\textit{Gender} (male = 1) & 0.271 (0.714) & -0.978** (0.484) & 1.540** (0.663) \\
\textit{Level of Studies} & 1.025** (0.512) & -0.092 (0.353) & 1.377*** (0.476) \\
\textit{GPA} & -0.594 (0.567) & 0.017 (0.378) & -0.753 (0.527) \\
\textit{Age} & -0.578 (0.462) & 0.526 (0.320) & -1.359*** (0.429) \\
\textit{Coding Experience} & 0.639 (0.606) & 0.276 (0.401) & 0.447 (0.563) \\
\textit{Python Experience} & -0.161 (0.803) & 0.119 (0.531) & -0.344 (0.747) \\
\textit{Studiousness} & -0.158 (0.259) & -0.066 (0.171) & -0.114 (0.241) \\
\textit{LLM Used Before} (yes = 1) & 0.537 (1.105) & 0.925 (0.730) & -0.478 (1.027) \\
\textit{LLM Experience} & -0.309 (0.290) & 0.081 (0.195) & -0.481* (0.270) \\
\textit{Pre-test} & 0.761*** (0.141) & 0.192* (0.106) & 0.701*** (0.131) \\
\textit{Learning Phase} &   & 0.812*** (0.073) &  \\
Constant & 5.994** (2.564) & -6.834*** (2.047) & 15.804*** (2.383) \\
\midrule \midrule
Observations & 107 & 107 & 107 \\
\(R^2\) & 0.472 & 0.772 & 0.512 \\
Adjusted \(R^2\) & 0.411 & 0.743 & 0.455 \\
\bottomrule
\end{tabular}}	
    {\textit{Notes}. Standard errors are in parenthesis. *: $p$<0.1; **: $p$<0.05; ***: $p$<0.01.}
\end{table}

%% file: methods.tex

In Study 1, we find a negative effect of substitutive LLM use on learning in the field. In Study 2, we further investigate the impact of LLMs on learning in an incentivized and pre-registered\endnotemark[\getrefnumber{endnote:prereg1}] laboratory experiment. We closely replicate the field study context by teaching experimental subjects Python programming in a manner similar to typical online courses (e.g., Udemy, Coursera). We manipulate the availability of an LLM and compare learning outcomes between subjects. The controlled setting in a laboratory allows us to ensure that subjects only use the provided materials for learning (e.g., no smartphones).

\subsection{Experimental Design}\label{sec:exp_design}

The main task of the subjects is to learn to code in Python. We choose Python as the programming language since it is among the most popular programming languages \citep{statista2024most} and is particularly beginner-friendly compared to other languages \citep{lemonaki2024most}.

The experiment consists of three distinct phases: pre-test, learning phase, and post-test. During the learning phase, which lasts for 45 minutes, subjects follow an introduction into basic Python programming, covering topics such as the print function, strings, and if-statements. The learning phase replicates the learning-by-doing approach of the assignments in Study 1. The learning phase is broken down into 24 parts. Each part comprises an explanation of a concept (e.g., if-statements), an example of how to implement it in Python, and a question in the form of a coding exercise for subjects to practice. The variable \textit{Learning Phase} counts the number of these practice questions that subjects answered correctly. Referring to our learning framework from Section \ref{sec:theory}, it measures the volume of topics that subjects were able to cover during the learning phase. To remove time effects, participants cannot continue to the post-test before the 45 minutes have elapsed.

The pre-test and post-test assess subjects' coding abilities before and after the learning phase. Both tests contain 20 programming questions and subjects have 20 minutes to solve them. The questions resemble those of the learning phase and cover the same topics. For each question in the pre-test, there exists a similar question in the  post-test, which requires knowledge of the same topic, such that we can directly compare the results of the two tests. The pre-test allows us to control for random differences in initial coding abilities between the control and treatment conditions and estimate learning in terms of "value added" \citep{hanushek2020education}. The variables \textit{Pre-test} and \textit{Post-test} count the number of questions that subjects answered correctly in the respective tests. Regarding learning outcomes, \textit{Post-test} measures students overall learning (i.e., through both topic volume and understanding).

During each of the three phases, subjects have access to a fully functional code editor and are able to submit code as often as they like. When submitting code, subjects immediately see their code's output and any potential error message. If subjects solve a question correctly (in the learning phase or in the tests), they see a short message confirming their result. Within each phase, subjects can move freely between the different questions.  We describe the user interface in detail in \autoref{sec:interface}.

Before the pre-test, we survey participants' demographics, studiousness, prior coding experience, and LLM experience (we describe all variables in \autoref{tab:covar_descr} in \autoref{sec:covariates}). After the post-test, we ask subjects to rate their perceived learning progress on a five-point Likert scale and to describe their learning experience in an open text field. All sessions took place in June 2024. The experiment was designed using oTree \citep{chen2016otree}.

\subsection{Treatment Condition}\label{sec:exp_groups}

We manipulate LLM availability by randomly assigning subjects to one of two conditions. In the control condition, subjects work through the experiment exactly as described above without access to an LLM. In the treatment condition, subjects have access to an LLM during the learning phase only (not in the pre or post-test). The LLM is available in a chat window next to the question description (see \autoref{fig:interface_treatment} in \autoref{sec:interface}). We use OpenAI's ChatGPT (\verb|gpt-3.5-turbo-0125|; \citealp{brown2020language}, \citealp{openai2024most}) as the LLM, which we query through an API. This model is trained to act as a helpful assistant and has sophisticated coding abilities \citep{liu2024your}. ChatGPT can solve all programming questions used in the experiment correctly when prompted with their descriptions. Subjects in the treatment condition can chat with the LLM as often as they like and with arbitrary message contents. They can reset the conversation with a button.

 \subsection{Subjects}\label{sec:subjects}
 
Subjects for the experiment were recruited from the laboratory participant pool at a public German university. The experiment was approved by the university's ethics committee.\endnote{Ethics Committee of the Faculty of Management, Economics and Social Sciences (ERC-FMES) at the University of Cologne. Reference number: 240020ML.} All participants are enrolled students and had no knowledge of the experimental content beforehand. Subjects were incentivized with a fixed compensation of 10 euros and a performance-based compensation depending on the number of solved questions in the post-test, where they received 1 euro per solved question. The incentive structure is in accordance with the laboratory's minimum wage requirements. 
 
We exclude participants based on two pre-registered filters. We exclude those who did not complete the entire experiment. In addition, we specifically designed the final question of the post-test such that is not solvable solely based on the knowledge acquired in the learning phase as it requires more advanced programming concepts. Thus, we use this question to filter out subjects with significant prior experience in Python as these are not the target group of our experiment.

%% file: revised_hypotheses.tex
\section{Revised Hypotheses}\label{sec:revised_hypotheses}

In Study 2, we do not find support for either Hypothesis \ref{hyp:hypo1} or Hypothesis \ref{hyp:hypo2}. While conducting the experiment, we noticed that the software on the computers in the laboratory blocked all options to copy and paste text. This mechanism was unintended by us and we hypothesize that this affects LLM usage and thus also learning outcomes.

Without copy and paste, subjects cannot easily copy task descriptions into the LLM or code snippets generated by the LLM back into the code editor. Instead, they have to manually copy text by typing it themselves. Such small increases in transaction costs are known to change behaviors \citep{lavecchia2016behavioral, ericson2019intertemporal}. Here, the inability to copy and paste poses a barrier to LLM usage and thereby also limits its effect on learning. For one, subjects may forgo the LLM entirely because they perceive the effort of manually copying questions and LLM-generated solutions as too high. Or, as we observed among several subjects in Study 2, they make mistakes and thus the LLM does not provide the correct answer, or they copy a solution incorrectly. Thus, LLM usage and consequentially its effect on learning are reduced without copy and paste. Consistent with behavioral research that incorporated initially unintended experimental conditions into their theorizing \citep[e.g.,][]{lee2024spaces}, we revise our hypotheses to include copy and paste:

\begin{rev_hypothesis}\label{hyp:rev_hypo_pro}
    When copy and paste are enabled, having access to LLMs while learning to code will increase the learning outcome of students.
\end{rev_hypothesis}

\begin{rev_hypothesis}\label{hyp:rev_hypo_contra}
    When copy and paste are enabled, having access to LLMs while learning to code will decrease the learning outcome of students.
\end{rev_hypothesis}

Further, we add a direct effect of copy and paste on LLM usage:

\begin{hypothesis}\label{hyp:rev_hypo_copy}
    When copy and paste are enabled, people use LLMs more during learning.
\end{hypothesis}


%% file: results.tex
\section{Study 3: Laboratory Experiment with Copy and Paste}\label{sec:s3}

To test the revised hypotheses, we conducted a second experiment with copy and paste.  Study 3 is an incentivized and pre-registered\endnote{\url{https://aspredicted.org/SXV_DLL}} laboratory experiment identical in its design to Study 2, except for the availability of copy and paste. Subjects could copy and paste text from question descriptions, code, and their conversations with the LLM (in the treatment condition) by right-clicking or with keyboard shortcuts. Beyond interacting with the LLM, the ability to copy and paste yielded no other discernible advantage in the experiment. All sessions took place in June 2024 three weeks after Study 2.

Of 72 enrolled participants, two were excluded for not completing the experiment and one for solving the final question (see Section \ref{sec:subjects}), resulting in 69 subjects. On average, subjects took 90 minutes and earned 17.88 euros. Prior Python coding experience was similar to Study 2 (78\% no prior experience and 14\% beginners). \autoref{tab:covar_stats} in \autoref{sec:covariates} provides summary statistics of all pre-treatment variables.

\autoref{tab:stud3_overview} shows the average performance of the treatment and control conditions in each of the experimental phases. The treatment condition outperforms the control condition in all three phases. In the \textit{Pre-test}, treated subjects solved one more question than controls ($p$ = 0.157); in the \textit{Learning Phase}, treated subjects solved four more questions ($p$ = 0.003); and in the \textit{Post-test}, treated subjects solved two more questions ($p$ = 0.092). The difference of \textit{Post-test} $-$ \textit{Pre-test} is on average one question larger for the treatment condition, but not statistically significant ($p$ = 0.312). In \autoref{tab:reg_s3_main}, the treatment effect on overall learning outcomes remains insignificant if we control for observed pre-treatment covariates (column 1, $p$ = 0.298). As in the model-free analysis, LLM access increased the volume of topics covered in the \textit{Learning Phase} (column 3, $p=0.013$), but it did not affect understanding when we control for subjects' progress during the \textit{Learning Phase} (column 2, $p$ = 0.359).\endnote{Results are similar if we exclude two subjects who did not use the LLM (see \autoref{tab:rob_posttest_treatment_excl_untreated} in \autoref{sec:robustness}).}

\begin{table}
    \TABLE
    {Study 3: Correctly Answered Questions per Experimental Phase.\label{tab:stud3_overview}}
	{\begin{tabular}{lcccc}
		\toprule
		     & \textit{Pre-test} & \textit{Learning Phase}  & \textit{Post-test} & \textit{Post-test} $-$ \textit{Pre-test} \\
		\midrule
		Control Condition    & 2.6 (2.4)  & 14.3 (5.0) & 6.8 (4.4) & 4.2 (3.5) \\
		Treatment Condition    & 3.7 (3.9) & 18.0 (4.8)  & 8.8 (5.1) & 5.0 (3.5) \\
		\bottomrule
	\end{tabular}}
    {\textit{Note}. Standard deviations are in parenthesis.}
\end{table}

\begin{table}
    \TABLE
    {Study 3: Regression Analyses.\label{tab:reg_s3_main}}
{\begin{tabular}{lccc}
\toprule
& \multicolumn{2}{c}{\textit{Post-test}} & \textit{Learning Phase} \\
& (1) & (2) & (3) \\
\midrule
\textit{Treatment} (LLM access = 1) & 0.959 (0.912) & -0.651 (0.705) & 2.614** (1.020) \\
\textit{Gender} (male = 1) & 1.919** (0.901) & 1.010 (0.672) & 1.474 (1.007) \\
\textit{Level of Studies} & 0.195 (0.587) & 0.410 (0.431) & -0.348 (0.656) \\
\textit{GPA} & -2.216** (0.834) & -0.489 (0.657) & -2.804*** (0.933) \\
\textit{Age} & -0.333 (0.376) & -0.283 (0.276) & -0.081 (0.421) \\
\textit{Coding Experience} & 1.780** (0.681) & 1.081** (0.508) & 1.134 (0.761) \\
\textit{Python Experience} & -0.676 (0.730) & -0.479 (0.535) & -0.319 (0.815) \\
\textit{Studiousness} & -0.262 (0.275) & -0.187 (0.201) & -0.120 (0.307) \\
\textit{LLM Used Before} (yes = 1) & 1.266 (1.273) & 0.923 (0.933) & 0.557 (1.423) \\
\textit{LLM Experience} & -0.333 (0.347) & -0.367 (0.254) & 0.056 (0.388) \\
\textit{Pre-test} & 0.769*** (0.145) & 0.325** (0.123) & 0.721*** (0.162) \\
\textit{Learning Phase} &   & 0.616*** (0.087) &  \\
Constant & 7.360** (3.085) & -2.987 (2.686) & 16.795*** (3.448) \\
\midrule \midrule
Observations & 69 & 69 & 69 \\
\(R^2\) & 0.638 & 0.810 & 0.600 \\
Adjusted \(R^2\) & 0.569 & 0.769 & 0.522 \\
\bottomrule
\end{tabular}}
    {\textit{Notes}. Standard errors are in parenthesis. *: $p$<0.1; **: $p$<0.05; ***: $p$<0.01.}
\end{table}

\section{Exploratory Analyses}\label{sec:exploratory}

To further investigate how LLMs affect learning, we conduct a number of exploratory analyses in the combined sample of Studies 2 and 3.\endnote{We mention the following exploratory analyses in both Studies' pre-registrations.} The studies are comparable because they were run within three weeks of each other in June 2024, both drew from the same subject pool and used the same lab, both followed the same experimental procedure, and the control conditions did not perform significantly different in any of the three experimental phases ($p=0.433$, $p=0.437$, $p=0.260$).

\subsection{Usage Behaviors}\label{sec:message_type_analysis}

In both experiments, we observe all messages subjects sent to the LLM. We manually code these messages based on what the subjects wanted the LLM to do. We distinguish between messages that asked for solutions (e.g., "complete the function hypotenuse...") and messages that asked for explanations (e.g., "explain what the str() function does"). When subjects asked for solutions, they substituted some of their learning activities (working through the practice exercises) with the LLM. In contrast, when subjects asked for explanations, they complemented their own learning activities with the LLM by requesting additional information on a subject. The two usage behaviors are similar to the output and input categories in \citet{keppler2024backwards}. Any other messages were either translations,\endnote{All translations are from English to German. We did not ask for subjects' nationality, but given the laboratory's location in Germany, the vast majority of subjects will have been German.} unrelated to the coding questions (miscellaneous), or user errors. See Appendix \ref{sec:message_coding} for details on the coding process and \autoref{tab:message_distribution} for the distributions of the usage behaviors. In our analyses, we focus on the number of times subjects substituted their learning activities by asking for \textit{Solutions} and the number of times they complemented their learning activities by asking for \textit{Explanations}.

\begin{table}
    \TABLE
    {Distributions of Usage Behaviors.\label{tab:message_distribution}}
	{\begin{tabular}{lccccc}
		\toprule
		     & \textit{Solutions} & \textit{Explanations}  & \textit{Miscellaneous} & \textit{Translations} & \textit{User Errors} \\
		\midrule
		Study 2    & 42.2\%  & 31.3\% & 11.6\% & 2.4\% & 12.4\% \\
		Study 3    & 64.2\% & 18.4\%  & 7.2\% & 6.1\% & 4.1\% \\
        Total & 54.1\% & 24.4\% & 9.2\% & 4.4\% & 7.9\% \\
		\bottomrule
	\end{tabular}}
    {}
\end{table}

We start by analyzing how the availability of copy and paste in Study 3 impacted how subjects used the LLM. Consistent with Hypothesis \ref{hyp:rev_hypo_copy}, copy and paste increased LLM use. On average, treated subjects in Study 3 sent 7.71 (S.D. = 5.50) messages to the LLM, which is significantly more than the 4.45 messages sent in Study 2 ($p$ = 0.002; \autoref{tab:message_type_determinants}, column 3: $p=0.001$). Only two subjects in Study 3 did not use the LLM at all. 

The availability of copy and paste not only changed how often subjects referred to the LLM, but also for which kinds of tasks they referred to it. As shown in columns 1 and 2 of \autoref{tab:message_type_determinants}, treated subjects in Study 3 asked for significantly more \textit{Solutions} ($p$ < 0.001) but not for more \textit{Explanations} ($p$ = 0.936) than treated subjects in Study 2. By making the LLM easier to use, copy and paste increased substitutive but not complementary use, suggesting that subjects intrinsically preferred the former. Furthermore, we note that prior experience with LLMs also determined usage behavior: subjects who had never used an LLM before (\textit{LLM Used Before} = 0) and those who used LLMs frequently (\textit{LLM Experience}) asked for \textit{Solutions} more often ($p=0.005$, $p=0.008$).

\begin{table}
    \TABLE
    {The Effect of Copy \& Paste on Usage Behaviors.\label{tab:message_type_determinants}}
{\begin{tabular}{lccc}
\toprule
& \textit{Solutions} & \textit{Explanations} & \textit{Messages} \\
& (1) & (2) & (3) \\
\midrule
\textit{Copy/Paste} (enabled = 1) & 3.066*** (0.755) & 0.032 (0.403) & 3.273*** (0.967) \\
\textit{Gender} (male = 1) & 1.337 (0.815) & -0.309 (0.435) & 1.007 (1.044) \\
\textit{Level of Studies} & -0.229 (0.547) & 0.273 (0.292) & -0.280 (0.701) \\
\textit{GPA} & 0.735 (0.643) & 0.028 (0.344) & 1.028 (0.824) \\
\textit{Age} & 0.300 (0.451) & -0.205 (0.241) & 0.299 (0.578) \\
\textit{Coding Experience} & -0.123 (0.706) & 0.778** (0.377) & 0.691 (0.904) \\
\textit{Python Experience} & 0.384 (0.728) & -0.692* (0.389) & -0.466 (0.933) \\
\textit{Studiousness} & -0.164 (0.279) & -0.148 (0.149) & -0.450 (0.358) \\
\textit{LLM Used Before} (yes = 1) & -3.538*** (1.228) & 0.581 (0.656) & -3.567** (1.573) \\
\textit{LLM Experience} & 0.890*** (0.328) & -0.138 (0.175) & 0.954** (0.421) \\
\textit{Pre-test} & -0.118 (0.126) & -0.005 (0.068) & -0.083 (0.162) \\
Constant & 1.079 (2.447) & 1.360 (1.307) & 4.713 (3.136) \\
\midrule \midrule
Observations & 94 & 94 & 94 \\
\(R^2\) & 0.322 & 0.084 & 0.264 \\
Adjusted \(R^2\) & 0.231 & -0.039 & 0.165 \\
\bottomrule
\end{tabular}}
    {\textit{Notes}. Regressions include treated subjects from Studies 2 and 3. Standard errors are in parenthesis. *: $p$<0.1; **: $p$<0.05; ***: $p$<0.01.}
\end{table}

We next use the availability of copy and paste in Study 3 to estimate how the associated change in usage behavior---specifically the increase in substitutive \textit{Solutions} requests---affected learning outcomes.\endnote{The availability of copy and paste did not affect learning outcomes other than through its effect on LLM usage behavior. The control condition in Study 3 (who could also use copy and paste) did not perform significantly different from the control condition in Study 2.} In \autoref{tab:reg_posttest_cp}, column 1, copy and paste did not affect overall learning outcomes ($p$ = 0.737). However, copy and paste decreased topic understanding (column 2, $p$ = 0.009) and increased topic volume (column 3, $p$  = 0.033). Treated subjects in Study 3 worked through 1.7 more question during the \textit{Learning Phase}, but holding this progress constant they solved 1.4 fewer questions in the \textit{Post-test}. Thus, when students use LLMs to substitute learning activities, there is a trade-off between volume and understanding. On the one hand, students can increase the volume of topics they cover because they are faster, which increases learning outcomes. On the other hand, this increase in volume comes at the cost of a lesser understanding of each topic, which decreases learning outcomes. Although treated subjects in Study 3 covered more topics than in Study 2, they understood each topic less. If we combine both effects, they cancel each other out, which explains the insignificant effect in column 1. The negative effect of substitution on understanding is likely due to reduced effort: In Study 3, 42\% of messages asking for a solution were sent without a single attempt to solve the corresponding question. Moreover, two participants remarked in the open text field at the end of the experiment that they did not learn anything as they relied too much on the LLM to solve the practice questions.

\begin{table}
    \TABLE
    {The Effect of Copy \& Paste on Learning Outcomes.\label{tab:reg_posttest_cp}}
{\begin{tabular}{lccc}
\toprule
& \multicolumn{2}{c}{\textit{Post-test}} & \textit{Learning Phase} \\
& (1) & (2) & (3) \\
\midrule
\textit{Copy/Paste} (enabled = 1) & -0.249 (0.737) & -1.407*** (0.525) & 1.733** (0.800) \\
\textit{Gender} (male = 1) & 2.143*** (0.796) & 0.549 (0.576) & 2.384*** (0.863) \\
\textit{Level of Studies} & 0.351 (0.534) & 0.352 (0.370) & -0.001 (0.580) \\
\textit{GPA} & -1.086* (0.629) & -0.259 (0.444) & -1.237* (0.682) \\
\textit{Age} & -0.297 (0.440) & -0.188 (0.305) & -0.163 (0.478) \\
\textit{Coding Experience} & 1.086 (0.690) & 1.027** (0.478) & 0.088 (0.748) \\
\textit{Python Experience} & -0.982 (0.712) & -0.646 (0.494) & -0.501 (0.772) \\
\textit{Studiousness} & 0.110 (0.273) & 0.067 (0.189) & 0.063 (0.296) \\
\textit{LLM Used Before} (yes = 1) & -0.365 (1.200) & 0.157 (0.832) & -0.780 (1.301) \\
\textit{LLM Experience} & -0.146 (0.321) & -0.253 (0.222) & 0.160 (0.348) \\
\textit{Pre-test} & 0.842*** (0.124) & 0.326*** (0.101) & 0.772*** (0.134) \\
\textit{Learning Phase} &   & 0.669*** (0.070) &  \\
Constant & 6.747*** (2.391) & -3.897* (2.000) & 15.913*** (2.594) \\
\midrule \midrule
Observations & 94 & 94 & 94 \\
\(R^2\) & 0.571 & 0.797 & 0.482 \\
Adjusted \(R^2\) & 0.513 & 0.767 & 0.413 \\
\bottomrule
\end{tabular}}
    {\textit{Notes}. Regressions include treated subjects from Studies 2 and 3. Standard errors are in parenthesis. *: $p$<0.1; **: $p$<0.05; ***: $p$<0.01.}
\end{table}

Lastly, to estimate how complementary use affected learning outcomes, we regress \textit{Post-test} on the number of \textit{Explanations} requested. Although we did not exogenously vary \textit{Explanations}, \textit{Pre-test} allows us to control for subjects' initial ability, which along with other pre-treatment observables is likely to be one of the main confounders. However, in this case we cannot estimate the effect of \textit{Explanations} on the volume of topics due to the simultaneity of usage behavior and \textit{Learning Phase} progress: Behavior affects progress but progress also affects behavior. In \autoref{tab:behavior_effects}, we see that \textit{Explanations} did not affect overall learning outcomes (column 1, $p=0.990$), but it did increase topic understanding (column 2, $p$ = 0.05). One reason for why the positive effect of \textit{Explanations} on understanding did not carry over to overall learning outcomes may be that asking for more \textit{Explanations} reduced topic volume. However, due to the aforementioned simultaneity, we cannot estimate this. We also include \textit{Solutions} in the regressions in \autoref{tab:behavior_effects} and the results are the same as with copy and paste in \autoref{tab:reg_posttest_cp}. In \autoref{sec:robustness}, we perform various robustness checks for this and the previous analyses and all results hold.

\begin{table}
    \TABLE
    {The Effect of Usage Behaviors on Learning Outcomes.\label{tab:behavior_effects}}
{\begin{tabular}{lcc}
\toprule
& \multicolumn{2}{c}{\textit{Post-test}} \\
& (1) & (2) \\
\midrule
\textit{Solutions} & -0.195* (0.107)  & -0.312*** (0.066) \\
\textit{Explanations} & -0.003 (0.200) & 0.249** (0.125) \\
\textit{Copy/Paste} (enabled = 1)  & 0.350 (0.802) & -0.574 (0.499) \\
\textit{Gender} (male = 1) & 2.404*** (0.805) & 0.885* (0.511) \\
\textit{Level of Studies} & 0.307 (0.533) & 0.213 (0.328) \\
\textit{GPA} & -0.943 (0.629) & 0.046 (0.395) \\
\textit{Age} & -0.239 (0.440) & -0.032 (0.271) \\
\textit{Coding Experience} & 1.064 (0.702) & 0.789* (0.432) \\
\textit{Python Experience} & -0.908 (0.721) & -0.321 (0.445) \\
\textit{Studiousness} & 0.077 (0.273) & 0.049 (0.167) \\
\textit{LLM Used Before} (yes = 1) & -1.055 (1.254) & -1.040 (0.770) \\
\textit{LLM Experience} & 0.027 (0.333) & 0.048 (0.205) \\
\textit{Pre-test} & 0.819*** (0.123) & 0.239** (0.091) \\
\textit{Learning Phase}  &  & 0.735*** (0.064) \\
Constant & 6.961*** (2.391) & -4.955*** (1.795) \\
\midrule \midrule
Observations & 94 & 94 \\
\(R^2\) & 0.588 & 0.847 \\
Adjusted \(R^2\) & 0.521 & 0.819 \\
\bottomrule
\end{tabular}}
    {\textit{Notes}. Regressions include treated subjects from Studies 2 and 3. Standard errors are in parenthesis. *: $p$<0.1; **: $p$<0.05; ***: $p$<0.01.}
\end{table}

How the different usage behaviors affect learning links back to our original hypotheses. As tutors, LLMs increase understanding if students ask for explanations and increase topic volume if students ask for solutions (Hypothesis \ref{hyp:rev_hypo_pro}). However, neither effect comes without a penalty. Asking for explanations does not increase overall learning outcomes, possibly because it reduces topic volume, while asking for solutions decreases understanding (Hypothesis \ref{hyp:rev_hypo_contra}). Therefore, whether LLMs help or hurt learning depends on how they are used and the context of the learning task. If topic volume matters more than understanding, using LLMs to speed up practice exercises may increase overall learning outcomes. If topic volume matters less than understanding (e.g., when students can learn without time constraints, as was the case in the field study), using LLMs to develop a deeper comprehension of topics may increase overall learning outcomes. Instead of either  Hypothesis \ref{hyp:rev_hypo_pro} or \ref{hyp:rev_hypo_contra} being true, our exploratory results suggest that they may both be true, depending on the way people use LLMs and the learning context.

\subsection{Heterogeneous Effects}

Previous studies have found that LLMs increase productivity the most for low-performing workers \citep{brynjolfsson2023generative, noy2023experimental, dell2023navigating}. Here, we test whether LLMs similarly affect learning depending on people's initial knowledge. To do so, we partition subjects in Studies 2 and 3 along the median of the combined sample's \textit{Pre-test} \citep[e.g.,][]{dell2023navigating}.

\autoref{tab:heteffects_pretest} shows the results for overall learning outcomes (column 1), understanding (column 2), and volume (column 3) in Studies 2 and 3 combined. The interaction of \textit{Treatment} and \textit{Pre-test $>$ Median} significantly affects understanding and overall learning outcomes (column 2: $p=0.034$; column 1: $p=0.028$). Treated subjects who scored higher on the \textit{Pre-test} increased their understanding and overall learning outcomes more than those who scored lower. Moreover, treated subjects who scored lower on the pre-test understood significantly less than control subjects (column 2: $p=0.043$). Topic volume is not affected by the interaction of \textit{Treatment} and \textit{Pre-test $>$ Median} ($p=0.290$). Unlike in workplaces, LLMs appear to increase inequality in education, especially with respect to understanding.

\begin{table}
    \TABLE
    {Heterogeneous Effects.\label{tab:heteffects_pretest}}
{\begin{tabular}{lccc}
\toprule
& \multicolumn{2}{c}{\textit{Post-test}} & \textit{Learning Phase} \\
& (1) & (2) & (3) \\
\midrule
\textit{Treatment} $\times$ \textit{Pre-test > Median} & 2.528** (1.142) & 1.613** (0.754) & 1.249 (1.177) \\
\textit{Treatment} (LLM access = 1) & 0.043 (0.701) & -0.950** (0.466) & 1.355* (0.722) \\
\textit{Copy/Paste} (enabled = 1) & -0.827 (0.553) & -1.222*** (0.365) & 0.539 (0.570) \\
\textit{Gender} (male = 1) & 0.872 (0.578) & -0.131 (0.386) & 1.368** (0.595) \\
\textit{Level of Studies} & 0.666* (0.390) & 0.212 (0.258) & 0.620 (0.402) \\
\textit{GPA} & -1.078** (0.466) & -0.004 (0.315) & -1.465*** (0.480) \\
\textit{Age} & -0.468 (0.306) & 0.037 (0.204) & -0.689** (0.316) \\
\textit{Coding Experience} & 1.083** (0.466) & 0.724** (0.307) & 0.490 (0.480) \\
\textit{Python Experience} & 0.296 (0.555) & 0.013 (0.366) & 0.387 (0.572) \\
\textit{Studiousness} & -0.188 (0.191) & -0.138 (0.126) & -0.068 (0.197) \\
\textit{LLM Used Before} (yes = 1) & 0.958 (0.833) & 0.798 (0.548) & 0.217 (0.858) \\
\textit{LLM Experience} & -0.169 (0.232) & -0.101 (0.153) & -0.092 (0.239) \\
\textit{Pre-test > Median} & 2.784*** (0.890) & 0.342 (0.609) & 3.333*** (0.917) \\
\textit{Learning Phase} &  & 0.733*** (0.050) &  \\
Constant & 6.544*** (1.958) & -4.541*** (1.494) & 15.128*** (2.017) \\
\midrule \midrule
Observations & 176 & 176 & 176 \\
\(R^2\) & 0.467 & 0.771 & 0.444 \\
Adjusted \(R^2\) & 0.424 & 0.751 & 0.399 \\
\bottomrule
\end{tabular}}
    {\textit{Notes}. Regressions include subjects from Studies 2 and 3. Standard errors are in parenthesis. *: $p$<0.1; **: $p$<0.05; ***: $p$<0.01.}
\end{table}

\subsection{Perceived learning}

Lastly, we investigate how LLMs affect perceptions of learning. Such perceptions can be different from actual learning and have been shown to disadvantage more effective teachers and teaching methods \citep{carrell2010does, deslauriers2019measuring}. Higher perceived learning affects choices \citep{jensen2010perceived} and in the case of LLMs may thus affect adoption. As we have shown above, such adoption is not always beneficial. LLMs increase perceptions of self-efficacy \citep{noy2023experimental} and may thus increase perceived learning.

At the end of the experiments, we asked subjects to rate their \textit{Perceived Learning} on a five-point Likert scale. \autoref{tab:reg_perceived} shows the results for the combined sample of Studies 2 and 3. In all regressions, we include the difference of subjects' \textit{Post-test $-$ Pre-test} to control for actual learning. The estimated \textit{Treatment} effect therefore represents subjects' perceptions that differ from actual learning. The effect of LLM exposure (\textit{Treatment}) on \textit{Perceived Learning} is positive and significant (column 1: $p=0.044$). This effect is independent of volume (column 2: $p=0.043$). Thus, LLMs increase perceived learning by more than can be explained by actual differences in learning.

\begin{table}
    \TABLE
    {The Effect of LLM Usage on Perceived Learning.\label{tab:reg_perceived}}
{\begin{tabular}{lcc}
\toprule
& \multicolumn{2}{c}{\textit{Perceived Learning}} \\
& (1) & (2) \\
\midrule
\textit{Treatment} (LLM access = 1) & 0.338** (0.166) & 0.348** (0.170) \\
\textit{Copy/Paste} (enabled = 1) & 0.128 (0.164) & 0.140 (0.170) \\
\textit{Post-test $-$ Pre-test} & 0.154*** (0.024) & 0.162*** (0.035) \\
\textit{Gender} (male = 1) & -0.079 (0.171) & -0.071 (0.174) \\
\textit{Level of Studies} & -0.079 (0.116) & -0.077 (0.117) \\
\textit{GPA} & -0.192 (0.140) & -0.198 (0.142) \\
\textit{Age} & -0.194** (0.091) & -0.197** (0.092) \\
\textit{Coding Experience} & 0.086 (0.138) & 0.083 (0.139) \\
\textit{Python Experience} & -0.101 (0.166) & -0.102 (0.166) \\
\textit{Studiousness} & 0.052 (0.057) & 0.053 (0.057) \\
\textit{LLM Used Before} (yes = 1) & -0.672*** (0.246) & -0.678*** (0.247) \\
\textit{LLM Experience} & 0.020 (0.068) & 0.020 (0.068) \\
\textit{Pre-test} & 0.083*** (0.030) & 0.092** (0.044) \\
\textit{Learning Phase} &  & -0.010 (0.035) \\
Constant & 2.975*** (0.591) & 3.079*** (0.699) \\
\midrule \midrule
Observations & 176 & 176 \\
\(R^2\) & 0.336 & 0.336 \\
Adjusted \(R^2\) & 0.283 & 0.278 \\
\bottomrule
\end{tabular}}
    {\textit{Notes}. Regressions include subjects from Studies 2 and 3. Standard errors are in parenthesis. *: $p$<0.1; **: $p$<0.05; ***: $p$<0.01.}
\end{table}

%% file: discussion.tex
\section{Discussion}\label{sec:discussion}

How do large language models (LLMs) affect student learning? We conceptualize learning outcomes as the value added by education \citep{hanushek2020education}, and decompose them into the volume of topics covered and the depth of understanding of each topic. In two pre-registered and incentivized laboratory experiments, we find no evidence that LLMs affect learning outcomes: neither the volume of topics that students go through nor the depth of their understanding of each topic changes when students have access to an LLM.

In exploratory analyses, we show that the effect of LLMs on learning outcomes depends on how students use them and their initial knowledge. Students who ask LLMs for solutions to practice exercises increase the number of topics they cover, but they understand each topic less. Students who ask LLMs for explanations of topics increase their understanding of each topic. In a field study of two programming course in which we only observe if students asked for solutions, we find a long-term and practically relevant negative effect of LLMs on learning. The ability to copy and paste when interacting with an LLM is a strong determinant of usage behavior as it substantially increases the number of times students prompt the LLM for solutions.

Unlike low-performing workers, who improve their productivity with LLMs more than high-performing workers \citep{brynjolfsson2023generative, dell2023navigating, noy2023experimental}, we find that students with less initial knowledge learn less when using LLMs. The unrestricted availability of LLMs thus increases inequality in learning outcomes. Moreover, students who have access to an LLM overestimate how much they have learned.

Our results have several theoretical implications for our understanding of how technology and LLMs in particular affect learning. First, artificial intelligence (AI) can substitute or complement human activities. Many of the reported productivity gains of AI have occurred in cases in which the technology complemented human work \citep{fugener2022cognitive, brynjolfsson2023generative, dell2023navigating, keppler2024backwards}. In education, both substitution (e.g., asking for solutions) and complementarity (e.g., asking for explanations) can support learning: Substitution increases the volume of knowledge that students learn because it speeds up the learning process, but it comes at the cost of reduced understanding. On the other hand, complementarity increases understanding. Our results indicate that in practical settings in which the time available for study is not a strongly limiting factor, understanding is more important than volume for overall learning outcomes. In such settings, LLMs should be used to complement students' learning activities rather than substitute them.

Second, previous debates on the efficacy of technology in education have focused on technology features (e.g., costs or interactivity) and how technology replaces other learning activities \citep{bulman2016technology, chatterji2018innovation}, but not on student behavior. Our results suggest that students prefer to use LLMs to substitute rather than complement learning activities. There are two possible explanations. For one, students may expect that substitution increases overall learning outcomes because it allows them to cover more topics. However, we do not see empirical evidence of improved actual or perceived learning outcomes from greater volume. A more likely explanation is that students' self-control problems extend to technology use. Learning is an effortful task and students exhibit a present bias which makes them prefer immediate gratification over exerting effort for learning in anticipation of future rewards \citep{lavecchia2016behavioral, damgaard2020behavioral}. Students may thus prefer substitution because it reduces their learning effort.

Third, a small change to the LLM interface can significantly alter student behavior. Specifically, the availability of copy and paste increases substitutive use but does not affect complementary use. Disabling copy and paste increases the immediate transaction costs of interacting with the LLM, thereby offsetting the decrease in learning effort.

Our study has a number of limitations that offer promising directions for future research. Despite the strong identification in our exploratory analyses, these findings should ideally be replicated in controlled experiments and with additional field data. Furthermore, we study learning outcomes that are a function of topic volume and topic understanding. We abstract away from the neurological building blocks of understanding, such as conceptual and factual understanding \citep{mueller2014pen} and memory \citep{wuttke2022seeing}. Studying how LLMs influence the neurological building blocks of learning may offer interesting insights. Moreover, LLMs may also affect learning in workplaces. As employees accumulate experience with a task, their performance tends to increase---so-called learning-by-doing \citep{arrow1962economic}. LLMs may change employee learning in two ways. On the positive side, LLMs can teach employees how to become better at their jobs, and this learning may persist without LLMs \citep[e.g.,][]{brynjolfsson2023generative}. On the negative side, LLMs may reduce how attentive employees work on a task or how much a task stimulates them mentally \citep[e.g.,][]{wang2020does, wuttke2022seeing, dell2022falling}. In the latter case, employees may not learn from experience anymore and may even unlearn some of their skills. As we have shown in education, the effect of LLMs on learning in workplaces likely depends on several factors, such as the complexity of the task involved and how employees use the LLM.

In practice, educators have been encouraged to embrace LLMs and mitigate their potential harm by advising students to use them as complementary aids for learning. However, with users' difficulty in maintaining self-control and the growing power of LLMs, adherence to voluntary guidelines may be elusive. This presents an ongoing challenge for education managers in both academic and corporate settings: How can they ensure that LLM-supported learning programs meet educational and training objectives while promoting high-quality knowledge transfer and equitable access for diverse learners.

Our experiments introduce a simple design tweak to a real-world LLM interface that fosters more effective knowledge transfer and can improve learning outcomes. This design may thus serve as a proof-of-concept for education managers seeking to implement LLM technology in their programs. We hope that its core principle---stimulating learners' mental engagement with rich problem-solving support while discouraging passive or superficial use---will inspire the development of innovative LLM-supported training programs that are effective in both academic and business environments.

%% file: screenshots.tex
\section{Experimental Interface}\label{sec:interface}

\begin{figure}[h]
\FIGURE
{\includegraphics[width=14cm]{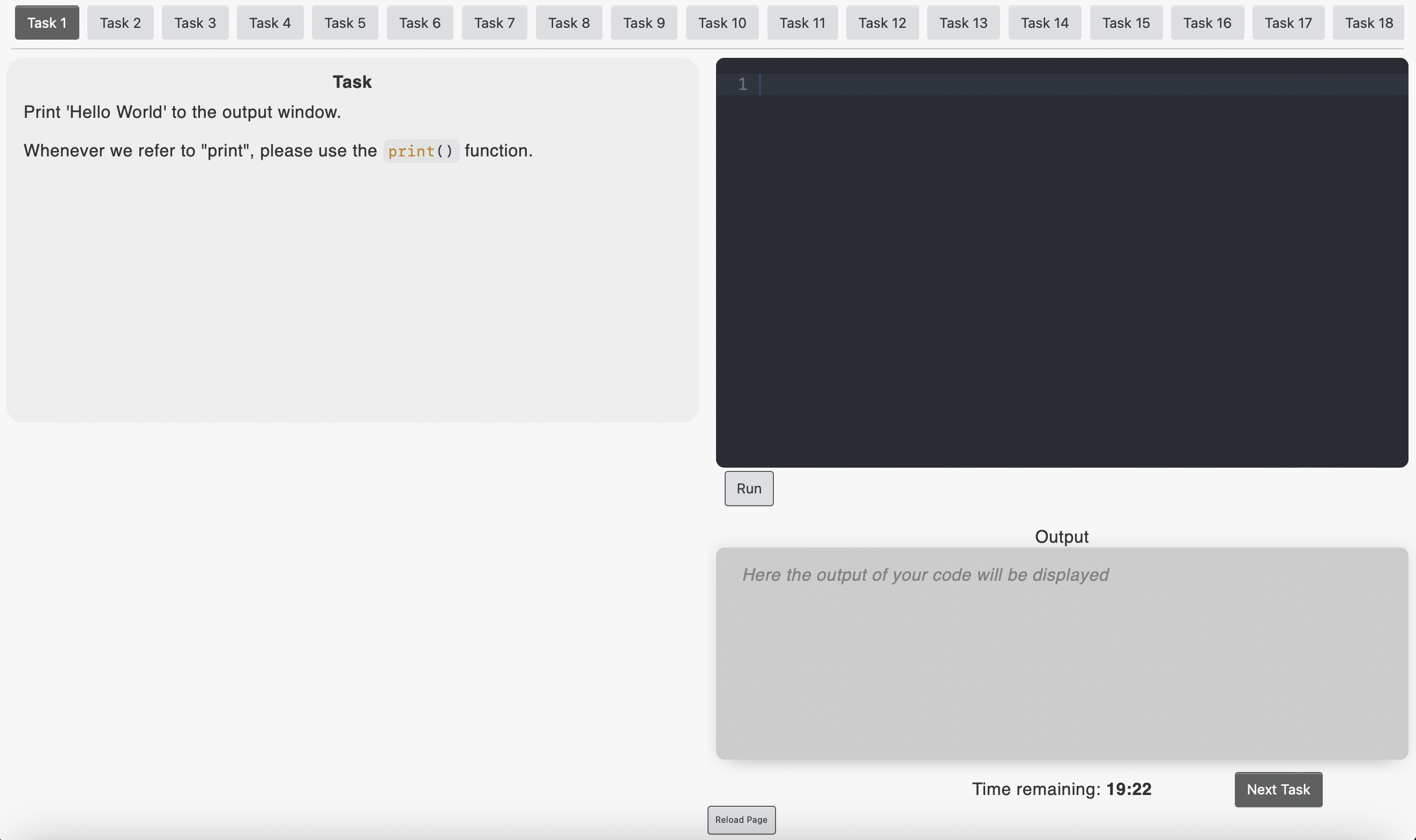}}
{Main experiment interface.\label{fig:interface_control}}
{}
\end{figure}

\begin{figure}[h]
\FIGURE
{\includegraphics[width=14cm]{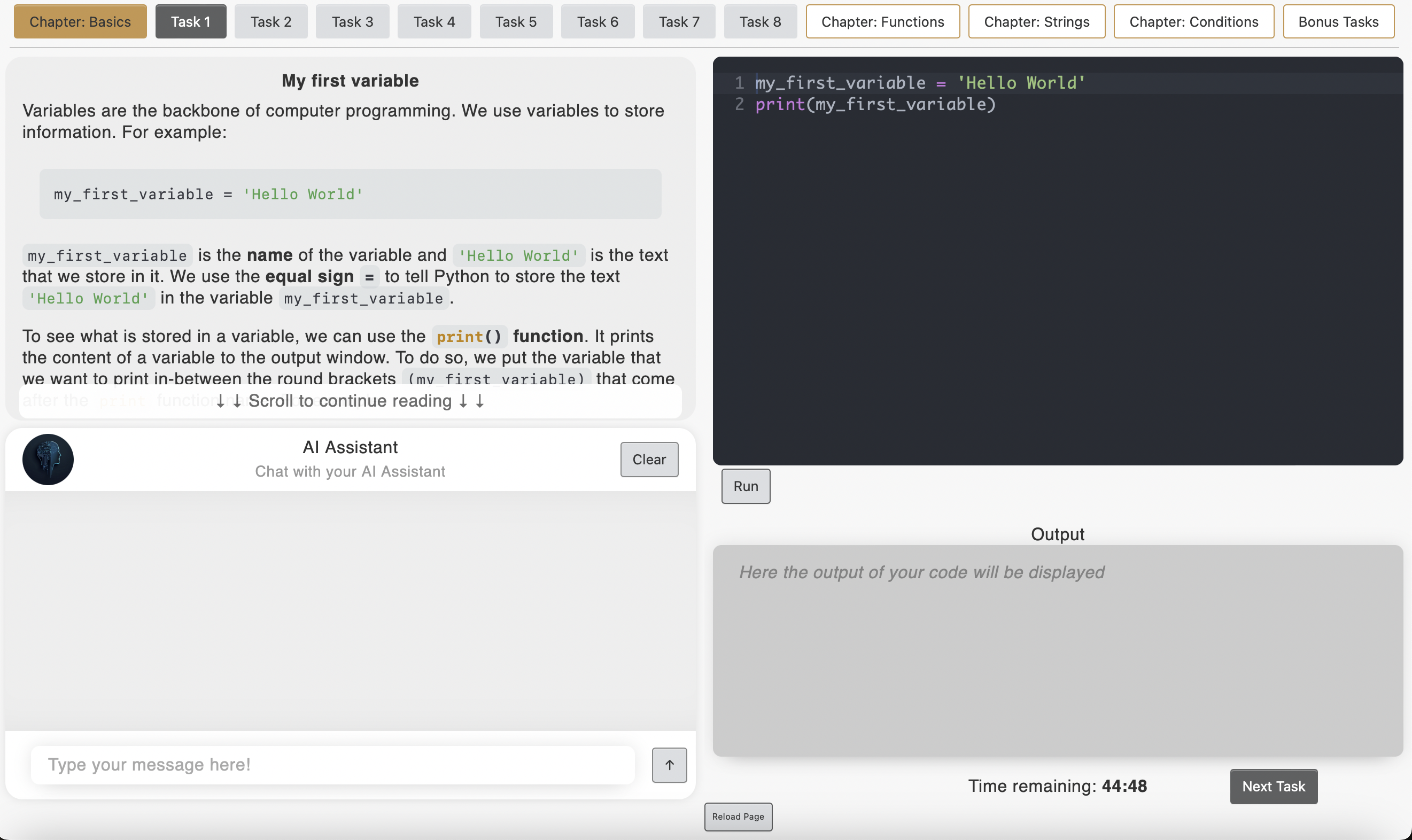}}
{Treatment Condition Interface.\label{fig:interface_treatment}}
{}
\end{figure}

Here, we describe the experimental interface with which subjects interacted in the pre-test, learning phase, and post-test. \autoref{fig:interface_control} shows a screenshot of the interface for the control condition during the pre-test. In the top left quadrant, subjects see the description of the current question and, in the learning phase, helpful information to learn how to solve this question. In the top right quadrant, subjects have access to a fully functioning code editor in which they can enter their Python code to solve the respective question. They can run code as often as they like. When running code, subjects see the full output below the editor including potential error messages and a short message if the question was solved correctly. We determine the correctness of solutions with predefined test cases. In addition to the above, the treatment condition also has access to an LLM during the learning phase in a chat window in the bottom left quadrant (see \autoref{fig:interface_treatment}).

%% file: covariates.tex
\section{Pre-treatment Variables}\label{sec:covariates}

In Studies 2 and 3, we measure a variety of pre-treatment variables. \autoref{tab:covar_descr} summarizes these variables and \autoref{tab:covar_stats} presents summary statistics.

\begin{table}
    \TABLE
    {Study 2 and 3: Description of Covariates.\label{tab:covar_descr}}
	{\begin{tabular}{c p{10cm}}
		\toprule
		Covariate & Description \\
		\midrule
		\textit{Gender}     & Indicator whether the participant is male. \\
        \textit{Level of Studies}     & The current level of studies, i.e. bachelor's, master's etc., of the participant.  \\
        \textit{GPA}		& Average grade of the participant on a German grading scale from 1 (best) to 4 (worst). \\
		\textit{Age}    & The age of the participant, discretized into buckets labeled from 1 to 5.  \\
		\textit{Coding Experience}     & Prior experience in any programming language on a five-point Likert scale.  \\
		\textit{Python Experience}     & Prior experience in Python on a five-point Likert scale. \\
		\textit{Studiousness}     & How much students learn on average for an exam on a five-point Likert scale.  \\
		\textit{LLM Used Before}     & Binary indicator whether the participant has used LLMs before.  \\
		\textit{LLM Experience}     & Experience with LLMs on a five-point Likert scale.  \\
		\bottomrule
	\end{tabular}}
    {}
\end{table}

\begin{table}
    \TABLE
    {Studies 2 and 3: Descriptive Statistics.\label{tab:covar_stats}}
	{\begin{tabular}{ccccc}
		\toprule
		& \multicolumn{2}{c}{Study 2} & \multicolumn{2}{c}{Study 3}                   \\
		\cmidrule(lr){2-3}\cmidrule(lr){4-5}
		     & Control & Treatment  & Control & Treatment \\
		\midrule
		\textit{Gender} (male = 1)     & 0.510  & 0.375 & 0.548 & 0.316 \\
		     &  (0.505) &  (0.489) &  (0.506) & (0.471) \\
        \textit{Level of Studies}     & 2.255  & 2.357  & 2.161  & 2.158  \\
		     &  (0.771) &  (0.750) &  (0.779) &  (0.718) \\
		\textit{GPA}		& 1.978  & 1.880  &  2.132  & 1.847  \\
				&  (0.570) &  (0.645) &   (0.664) &  (0.473) \\
		\textit{Age}    & 2.569 & 2.643  & 2.839  & 2.395  \\
		    &  (0.878) &  (0.796) &  (1.241) &  (1.054) \\
		\textit{Coding Experience}     & 1.471  & 1.767  & 1.806  & 1.711  \\
		     &  (0.674) &  (0.809) &  (0.910) &  (0.732) \\
		\textit{Python Experience}     & 1.176  & 1.339  & 1.290  & 1.342   \\
		     & (0.434) &  (0.611) &  (0.588) &  (0.781)  \\
		\textit{Studiousness}     & 4.373 & 4.375 & 4.000  & 4.105   \\
		     &  (1.312) &  (1.342) &  (1.789) &  (1.448)  \\
		\textit{LLM Used Before} (yes = 1)     & 0.804  & 0.911  & 0.871  & 0.816  \\
		     &  (0.401) &  (0.288) &  (0.341) &  (0.393) \\
		\textit{LLM Experience}     & 2.353  & 2.321  & 2.387  & 2.000   \\
		     &  (1.415) &  (1.336) &  (1.476) &  (1.252)  \\
		\bottomrule
	\end{tabular}}
    {\textit{Note.} We report the means and in brackets the standard deviations.}
\end{table}

%% file: message_coding.tex
\section{Coding of Chat Messages}\label{sec:message_coding}

We code each message sent to the LLM. We blindly assess the messages and assign them to a category based on the intention of the subject. Categories were created inductively during the coding process whenever we found a message that did not fit into a previously created category. \autoref{tab:message_coding} summarizes the resulting categories of this coding procedure. The categories are clearly distinct from each other. \textit{Solutions} are easily identifiable by containing phrases such as "solve this task" or the copied question description. \textit{Explanations} are identifiable by asking other problem related questions without explicitly asking for a solution. 

\begin{table}
    \TABLE
    {Chat Messages Types.\label{tab:message_coding}}
	{\begin{tabular}{l p{7cm} p{5cm} }
		\toprule
		Category & Description & Example \\
		\midrule
		\textit{Solutions}     & User asks the LLM to solve an exercise, most commonly by pasting in the entire task description. & \textit{"Complete the function hypotenuse..."}  \\
		\textit{Explanations}     & User asks the LLM to explain a programming concept.  & \textit{"Explain what the str() function does"} \\
		\textit{Miscalleneous} 		& Messages not related to the tasks. & \textit{"hows your day mate?"} \\
		\textit{Translations}    & User asks for a translation from English to German.  & \textit{"what is the german word for concatenate"} \\
		\textit{User Errors}     & Most commonly users ask questions without providing context and assume the LLM has access to everything onscreen. & \textit{"What am I doing wrong here?"}  \\
		\bottomrule
	\end{tabular}}	
    {}
\end{table}

%% file: robustness.tex
\section{Robustness Checks}\label{sec:robustness}

In this section, we provide additional robustness checks of our results.

First, our primary analyses focus on intention-to-treat effects since subjects in the treatment group only have access to the LLM but must not necessarily use it. We repeat our analyses on the effects of treatment in Study 2 (\autoref{tab:reg_s2_main}) and Study 3 (\autoref{tab:reg_s3_main}) in \autoref{tab:rob_posttest_treatment_excl_untreated} after excluding unsuccessfully treated subjects, i.e. those who did not use the LLM at all (six in Study 2 and two in Study 3). The results conform to our previous findings as we find no effect of LLM usage (\textit{Treatment}) on learning outcomes (\textit{Post-test}, column 1: $p=0.868$, column 4: $p=0.333$) or on understanding (column 2: $p=0.319$, column 5: $p=0.342$) and no effect on topic volume in Study 2 (column 3, $p=0.262$) but a positive effect in Study 3 (column 6, $p=0.015$).

\begin{rotate}
\begin{table}[p]
    \TABLE
    {Studies 2 and 3: Treatment Effects Without Unsuccessfully Treated.\label{tab:rob_posttest_treatment_excl_untreated}}
{\small\begin{tabular}{lcccccc}
\toprule
& \multicolumn{3}{c}{Study 2} & \multicolumn{3}{c}{Study 3} \\
\cmidrule(lr){2-4}\cmidrule(lr){5-7}
& \multicolumn{2}{c}{\textit{Post-test}} & \textit{Learning Phase} & \multicolumn{2}{c}{\textit{Post-test}} & \textit{Learning Phase} \\
& (1) & (2) & (3) & (4) & (5) & (6) \\
\midrule
\textit{Treatment} (LLM access = 1) & 0.122 (0.733) & -0.495 (0.494) & 0.760 (0.674) & 0.920 (0.941) & -0.701 (0.731) & 2.623** (1.043) \\
\textit{Gender} (male = 1) & 0.274 (0.733) & -0.986* (0.505) & 1.553** (0.674) & 1.936** (0.913)	 & 1.019 (0.685) & 1.484 (1.011) \\
\textit{Level of Studies} & 1.129** (0.530) & -0.072 (0.373) & 1.479*** (0.487) & 0.193 (0.597) & 0.425 (0.440) & -0.375 (0.661) \\
\textit{GPA} & -0.443 (0.573) & 0.106 (0.387) & -0.676 (0.526) & -2.234** (0.846) & -0.495 (0.671) & -2.814*** (0.937) \\
\textit{Age} & -0.439 (0.486) & 0.600* (0.340) & -1.280*** (0.447) & -0.305 (0.389) & -0.301 (0.286) & -0.006 (0.431) \\
\textit{Coding Experience} & 0.621 (0.609) & 0.334 (0.408) & 0.354 (0.559) & 1.807** (0.690) & 1.086** (0.519) & 1.166 (0.765) \\
\textit{Python Experience} & -0.266 (0.812) & 0.066 (0.545) & -0.409 (0.746) & -0.682 (0.738) & -0.481 (0.544) & -0.325 (0.818) \\
\textit{Studiousness} & -0.176 (0.267) & -0.099 (0.179) & -0.094 (0.245) & -0.264 (0.278) & -0.188 (0.205) & -0.124 (0.308) \\
\textit{LLM Used Before} (yes = 1) & 0.470 (1.117) & 0.748 (0.749) & -0.342 (1.027) & 1.245 (1.309) & 0.864 (0.965) & 0.617 (1.450) \\
\textit{LLM Experience} & -0.280 (0.293) & 0.145 (0.201) & -0.523* (0.270) & -0.342 (0.351) & -0.364 (0.259) & 0.035 (0.389) \\
\textit{Pre-test} & 0.788*** (0.142) & 0.211* (0.110) & 0.710*** (0.131) & 0.773*** (0.148) & 0.329** (0.127) & 0.718*** (0.164) \\
\textit{Learning Phase} &   & 0.812*** (0.077) &  &  & 0.618*** (0.090) &  \\
Constant& 5.233* (2.662) & -7.192*** (2.138) & 15.301*** (2.446) & 7.317** (3.131) & -2.957 (2.743) & 16.622*** (3.467) \\
\midrule \midrule
Observations & 101 & 101 & 101 & 67 & 67 & 67 \\
\(R^2\) & 0.484 & 0.771 & 0.523 & 0.637 & 0.807 & 0.604 \\
Adjusted \(R^2\) & 0.420 & 0.740 & 0.464 & 0.565 & 0.765 & 0.525 \\
\bottomrule
\end{tabular}}
    {\textit{Notes}. We exclude subjects in the treatment condition who did not use the LLM at all. Standard errors are in parenthesis. *: $p$<0.1; **: $p$<0.05; ***: $p$<0.01.}
\end{table}
\end{rotate}

Second, we replicate the analyses in \autoref{tab:behavior_effects} (Section \ref{sec:exploratory}) with a relative measure of usage behavior by including the total number of \textit{Messages} sent to the LLM as an additional control. The results are shown in \autoref{tab:rob_behavior_messages}. Including this control subsumes the effect of \textit{Solutions} across all specifications ($p=0.906$, $p=0.090$). However, we remark that \textit{Messages} is highly correlated with \textit{Solutions} with a Pearson correlation coefficient of $\rho = 0.81$ ($p<0.001$) and produces a Variable Inflation Factor greater than ten, indicating significant multicollinearity and providing grounds for exclusion \citep{o2007caution}. 

\begin{table}
    \TABLE
    {Studies 2 and 3: Additional Analyses of the Effect of Usage Behaviors on Learning Outcomes.\label{tab:rob_behavior_messages}}
{\begin{tabular}{lcc}
\toprule
& \multicolumn{2}{c}{\textit{Post-test}} \\
& (1) & (2) \\
\midrule
\textit{Solutions} &  0.025 (0.207) & -0.223* (0.130) \\
\textit{Explanations} & 0.264 (0.294) & 0.355* (0.182) \\
\textit{Copy/Paste} (enabled = 1) & 0.415 (0.801) & -0.541 (0.501) \\
\textit{Gender} (male = 1) & 2.422*** (0.802) & 0.903* (0.513) \\
\textit{Level of Studies} & 0.221 (0.536) & 0.179 (0.331) \\
\textit{GPA} & -0.877 (0.629) & 0.065 (0.397) \\
\textit{Age} & -0.182 (0.441) & -0.011 (0.273) \\
\textit{Coding Experience} & 1.041 (0.700) & 0.782* (0.433) \\
\textit{Python Experience} & -0.914 (0.718) & -0.328 (0.446) \\
\textit{Studiousness} & 0.050 (0.273) & 0.038 (0.168) \\
\textit{LLM Used Before} (yes = 1) & -1.246 (1.259) & -1.117 (0.778) \\
\textit{LLM Experience} & 0.086 (0.336) & 0.072 (0.207) \\
\textit{Pre-test} & 0.828*** (0.123) & 0.246*** (0.092) \\
\textit{Learning Phase} &  & 0.730*** (0.064) \\
\textit{Messages} & -0.228 (0.184) & -0.092 (0.115) \\
Constant & 7.437*** (2.414) & -4.677** (1.832) \\
\midrule \midrule
Observations & 94 & 94 \\
\(R^2\) & 0.596 & 0.848 \\
Adjusted \(R^2\) & 0.524 & 0.819 \\
\bottomrule
\end{tabular}}
    {\textit{Notes}. Regressions only include treated subjects. Standard errors are in parenthesis. *: $p$<0.1; **: $p$<0.05; ***: $p$<0.01.}
\end{table}

Finally, we also consider an alternative dependent variable to measure understanding. Previously, we inferred understanding by measuring learning outcomes (\textit{Post-test}) while controlling for volume (\textit{Learning Phase}). In \autoref{tab:rob_understanding_new_dv}, we replicate our main analyses on understanding with another variable. \textit{Covered Post-test} measures the fraction of \textit{Post-test} questions solved whose topics the subject covered during the learning phase. In other words, \textit{Covered Post-test} is normalized by learning phase progress. As in \autoref{tab:reg_s2_main} and \autoref{tab:reg_s3_main}, LLM access has no effect on understanding (columns 1 and 2, $p=0.527$, $p=0.580$). Also similarly to our earlier analyses in \autoref{tab:reg_posttest_cp} and \autoref{tab:behavior_effects} in Section \ref{sec:exploratory}, we find a negative (albeit now weaker) effect of copy and paste on understanding ($p=0.086$). This effect vanishes once we include \textit{Solutions} and \textit{Explanations} ($p=0.458$). Conforming to our main analysis, \textit{Solutions} decrease \textit{Covered Post-test} ($p=0.018$) whereas \textit{Explanations} increase \textit{Covered Post-test} ($p=0.003$).

\begin{table}
    \TABLE
    {Additional Analyses of the Effects on Understanding.\label{tab:rob_understanding_new_dv}}
{\begin{tabular}{lcccc}
\toprule
& \multicolumn{4}{c}{Solved of Covered} \\
\cmidrule(lr){2-5}
& Study 2 & Study 3 & \multicolumn{2}{c}{Studies 2 \& 3} \\
& (1) & (2) & (3) & (4) \\
\midrule
\textit{Solutions} &  &  &   & -0.016** (0.007) \\
\textit{Explanations} &  &  &   & 0.038*** (0.012) \\
\textit{Treatment} (LLM access = 1) & -0.034 (0.054) & -0.036 (0.065) &  &  \\
\textit{Copy/Paste (enabled = 1)} &  &  & -0.085* (0.049) & -0.037 (0.050) \\
\textit{Gender (male = 1)} & -0.127** (0.054) & 0.024 (0.064) & 0.001 (0.053) & 0.034 (0.050) \\
\textit{Level of Studies} & 0.067* (0.039) & 0.002 (0.042) & 0.086** (0.035) & 0.071** (0.033) \\
\textit{GPA} & -0.010 (0.043) & -0.064 (0.059) & -0.036 (0.042) & -0.025 (0.039) \\
\textit{Age} & -0.010 (0.035) & -0.060** (0.027) & -0.064** (0.029) & -0.051* (0.027) \\
\textit{Coding Experience} & 0.045 (0.046) & 0.061 (0.048) & 0.115** (0.046) & 0.084* (0.043) \\
\textit{Python Experience} & -0.039 (0.061) & -0.001 (0.052) & -0.070 (0.047) & -0.038 (0.044) \\
\textit{Studiousness} & 0.011 (0.020) & -0.039** (0.019) & 0.017 (0.018) & 0.020 (0.017) \\
\textit{LLM Used Before (yes = 1)} & 0.190** (0.084) & 0.187** (0.090) & 0.109 (0.079) & 0.030 (0.077) \\
\textit{LLM Experience} & -0.019 (0.022) & -0.020 (0.025) & -0.033 (0.021) & -0.013 (0.021) \\
\textit{Pre-test} & 0.024** (0.011) & 0.026** (0.010) & 0.022** (0.008) & 0.020** (0.008) \\
Constant & 0.378* (0.194) & 0.740*** (0.218) & 0.451*** (0.158) & 0.417*** (0.148) \\
\midrule \midrule
Observations & 107 & 69 & 94 & 94 \\
\(R^2\) & 0.222 & 0.403 & 0.361 & 0.465 \\
Adjusted \(R^2\) & 0.132 & 0.287 & 0.276 & 0.378 \\
\bottomrule
\end{tabular}}
    {\textit{Notes}. Columns 1 and 2 include all subjects, columns 3 and 4 include only treated subjects. Standard errors are in parenthesis. *: $p$<0.1; **: $p$<0.05; ***: $p$<0.01.}
\end{table}